\documentclass[english,aps,prl,twocolumn,amsmath,amssymb,showpacs,superscriptaddress,notitlepage,longbibliography]{revtex4-2}
\usepackage[T1]{fontenc}
\usepackage[latin9]{inputenc}
\usepackage{color}
\usepackage{mathtools}
\usepackage{dsfont}
\usepackage{amsmath}
\usepackage{amssymb}
\usepackage{graphicx}

\usepackage{pdfpages}
\makeatletter
\AtBeginDocument{\let\LS@rot\@undefined}
\makeatother

\makeatletter

\usepackage{amssymb}
\usepackage{amsmath}
\usepackage{graphicx}
\usepackage[colorlinks=true,linkcolor=blue,anchorcolor=red,citecolor=blue,urlcolor=blue]{hyperref}
\usepackage[caption=false]{subfig}
\usepackage{lipsum}

\DeclareMathOperator{\sgn}{sgn}
\DeclareMathOperator{\pf}{Pf}
\DeclareMathOperator{\antidiag}{anti-diag}

\makeatother

\usepackage{babel}
\begin{document}
\renewcommand{\figurename}{Fig.}
\title{Making topologically trivial non-Hermitian systems nontrivial via gauge fields}
\author{W.\ B. Rui}
\email{wbrui@hku.hk}

\address{Department of Physics and HK Institute of Quantum Science \& Technology, The University of Hong Kong, Pokfulam Road, Hong Kong, China}

\author{Y.\ X. Zhao}
\email{yuxinphy@hku.hk}

\address{Department of Physics and HK Institute of Quantum Science \& Technology, The University of Hong Kong, Pokfulam Road, Hong Kong, China}

\author{Z.\ D. Wang}
\email{zwang@hku.hk}

\address{Department of Physics and HK Institute of Quantum Science \& Technology, The University of Hong Kong, Pokfulam Road, Hong Kong, China}

\begin{abstract}
Non-Hermiticity significantly enriches the concepts of symmetry and topology in physics. Particularly, non-Hermiticity gives rise to the ramified symmetries, where the non-Hermitian Hamiltonian $H$ is transformed to $H^\dagger$. For time-reversal ($T$) and sublattice symmetries, there are six ramified symmetry classes leading to novel topological classifications with various non-Hermitian skin effects. As artificial crystals are the main experimental platforms for non-Hermitian physics, there exists the symmetry barrier for realizing topological physics in the six ramified symmetry classes: While artificial crystals are in spinless classes with $T^2=1$, nontrivial classifications dominantly appear in spinful classes with $T^2=-1$. Here, we present a general mechanism to cross the symmetry barrier. With an internal parity symmetry $P$, the square of the combination $\tilde{T}=PT$ can be modified by appropriate gauge fluxes. Using the general mechanism, we systematically construct spinless models for all non-Hermitian spinful topological phases in one and two dimensions, which are experimentally realizable. Our work suggests that gauge structures may significantly enrich non-Hermitian physics at the fundamental level.
\end{abstract}
\maketitle
\textit{\textcolor{blue}{Introduction.}}\textit{\textemdash{}}Recently, non-Hermitian topology has attracted great interests  \citep{LeePRL2016,Leykam_PRL_2017,shen_PRL_2018,zhou_observation_2018,KunstPRL2018,rui_classification_2019,LeePRL2019,song2019nonhermitian,Rui_PT_2019,LeePRLAnomalous2019,YoshidaPRB2019,ashidaNH2020,BorgniaPRL2020,liCriticalNonHermitianSkin2020,LonghiPRL2019,BergholtzRMP2021,BitanPRB2022,ruiHermitian2022,delplace2021symmetryprotected,rui_spatial_2022},
ranging from fundamental concepts to fascinating phenomena, such as
complex-energy gaps \citep{gong2018,kawabata2019symmetry}, 
non-Bloch band theory \citep{WangPRL2018,yokomizo2019nonbloch,yang2020nonhermitian,kawabata2020nonbloch},
exceptional degeneracies \citep{miriEP2019,kawabata2019classification,kato_perturbation,heiss2012thephysics,berry2004physics,Xu_weyl_nh},
and non-Hermitian skin effects \citep{WangPRL2018,okuma2020topological,zhang2020correspondence,HelbigNH2020,ghatak2020observation,zhang2022universal}. Analogous to the Hermitian topology, symmetry remains to be foundational for non-Hermitian
topology, as it provides the framework for topological classifications
\citep{kawabata2019symmetry,ZhouNHclassify}. But for a non-Hermitian Hamiltonian $H$, the Hermitian conjugate $H^\dagger$ is not equal to $H$. This gives rise to a new possibility for symmetry. A symmetry may transform $H$ to $H^\dagger$, rather than leave $H$ invariant. Such a symmetry is referred to as ramified symmetry \citep{kawabata2019symmetry,ZhouNHclassify,bernard_classification_2002,Lieu2018}. 
As a prominent example, the ten Altland-Zirnbauer (AZ) symmetry classes have been extended into 38 symmetry classes with eight additional ramified AZ$^\dagger$ symmetry classes~\cite{footnote}. Topological classifications have been worked out for the ramified symmetry classes; nontrivial topology can lead to the non-Hermitian skin effect, featuring the breakdown of conventional bulk-boundary correspondence.

\begin{figure*}[t]
	\includegraphics[width=\textwidth]{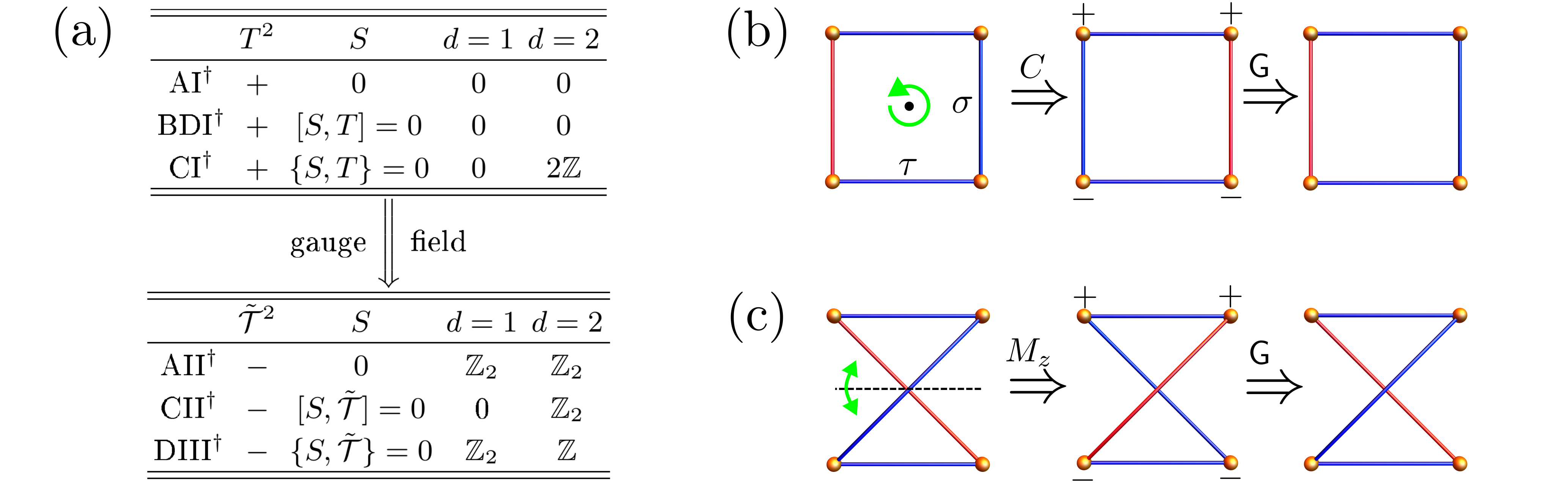}
	\caption{(a) Classification of spinless (upper table) and spinful (lower table) classes with ramified time-reversal and sublattice symmetries for point-gap topological phases~\cite{gapcondition,kawabata2019symmetry}. See Ref.~\cite{,commutation} for the connection of our convention to the previous convention of the symmetry classes. The symmetry barrier between spinless and spinful classes can be crossed by gauge field. (b) Rectangle with $\pi$ gauge flux, invariant under the projective twofold rotation. (c) Twisted rectangle with $\pi$ gauge flux, invariant under the projective mirror reflection. Blue and red denote the $+$ and $-$ hopping amplitudes, respectively. The gauge transformations ($\mathsf{G}$) are specified by the signs in the two middle panels. \label{fig:Fig1}}
\end{figure*}

Here, among the eight ramified AZ$^\dagger$ symmetry classes, we focus on the six ones with time reversal ($T$) symmetry. Note that the other two contain particle-hole symmetry for superconductors, and therefore are not considered here. We further categorize the six classes with $T$ symmetry into three spinless and three spinful classes with $T^2=1$ and $T^2=-1$, respectively. The topological classification table for point-gap systems~\cite{gapcondition} in one and two dimensions is given by Fig.~\ref{fig:Fig1}(a)~\cite{commutation}, as the skin effect is mainly investigated in the low dimensions. From Fig.~\ref{fig:Fig1}(a), it is exciting to observe various nontrivial topological classifications, which dominantly belong to the spinful classes. The situation is unfortunate for experimental realization, because non-Hermitian topology is mainly realized by artificial crystals, e.g., photonic/acoustic crystals and electric-circuit arrays. All these artificial crystals satisfy $T^2=1$, and therefore belong to spinless classes. Hence, we see that there is the symmetry barrier preventing broad realizations of non-Hermitian topological phases by artificial crystals.

In this Letter, we provide general mechanisms, by which the symmetry barrier can be crossed. The key observation is that for $1$D and $2$D systems, we can consider two anti-unitary symmetries, $T$ and $\tilde{T}=PT$, with $P$ a twofold internal symmetry. For both $1$D and $2$D systems, $P$ can be the mirror reflection $M_z$ inversing the $z$ axis, presuming that the system is placed on the $x$-$y$ plane. For a $1$D system, we may choose $P$ as the twofold rotation $C$ through the direction of the $1$D system.

\begin{figure*}[t]
	\includegraphics[width=0.9\textwidth]{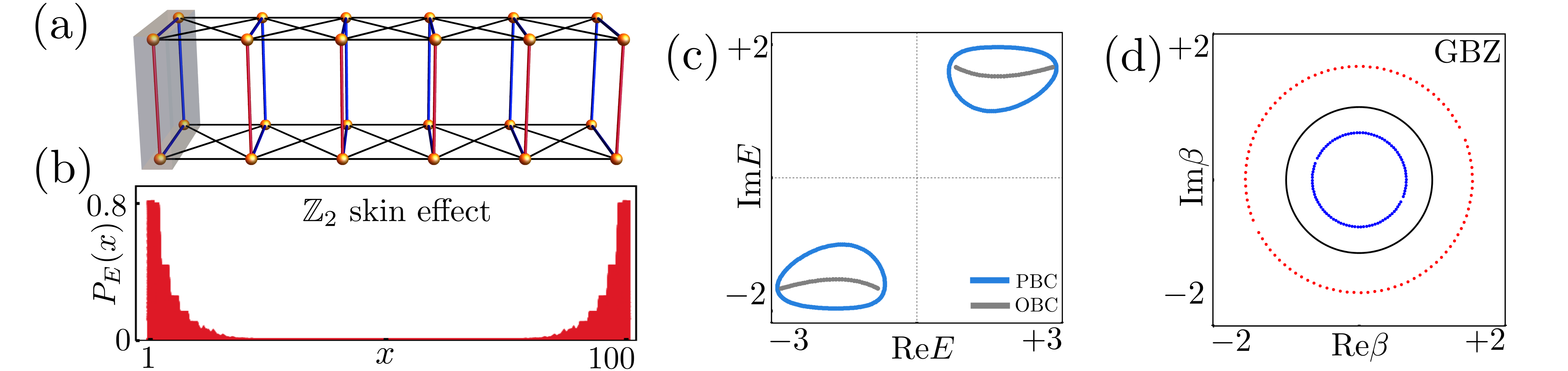}
	
	\caption{(a) Schematic picture for the 1D spinless chain. The unit cell is constructed by the rectangle in Fig.~\ref{fig:Fig1}(b), which is marked by gray. (b) The $\mathbb{Z}_2$ non-Hermitian skin effect.   For the Hamiltonian with 100 unit cells under the open boundary conditions, we compute all the energy eigenstates $|\psi_E\rangle$ with energy $E$. For each $E$, we plot $P_E(x)=\sum_\alpha |\langle x,\alpha|\psi_E \rangle|^2$, where $|x,\alpha\rangle$ is the on-site state at $\alpha$th site in the $x$th unit cell. (c) The energy spectra and (d) the generalized Brillouin zone (blue and red) of the system. The black circle in (d) denotes $|\beta|=1$. The parameters are $J_{R}=J_{I}=1.5,t=\mu=1,\gamma=2.$
		\label{fig:Fig2}}
\end{figure*}

Ordinarily, $\tilde{T}^2=T^2$, but by turning on in-plane or in-line gauge fluxes, we may have
\begin{equation}\label{eq:Switch_T}
	\tilde{\mathcal{T}}^2=-T^2,
\end{equation} 
Here, the font is changed with gauge fluxes, noting that the intrinsic $T$ cannot be changed by gauge fluxes. This is because the symmetry group is projectively represented in the presence of gauge fluxes, i.e., gauge fluxes can induce additional phase factors into the symmetry multiplications, resembling the Aharonov-Bohm effect. Particularly, for $P=C$ or $M_z$ the additional minus sign in \eqref{eq:Switch_T} shall be realized with appropriate gauge fluxes.  Then, for spinless systems, we can realize $\tilde{\mathcal{T}}^2=-T^2=-1$ according to \eqref{eq:Switch_T}. Thus, $\tilde{\mathcal{T}}$ becomes the fundamental symmetry for us. To realize spinful non-Hermitian topologies by spinless systems, we can preserve $\tilde{\mathcal{T}}$ while allowing $T$ to be broken. This is illustrated in Fig.~\ref{fig:Fig2}(b); with the insertion of gauge fluxes, the spinful skin states rise on a previously trivial spinless chain.

We thoroughly implement the mechanism for all nontrivial cases in the classification table of Fig.~\ref{fig:Fig1}(a). That is, we systematically construct spinless models that can realize all spinful non-Hermitian topological phases. All the spinless models for spinful non-Hermitian topological phases can be realized by artificial crystals. For demonstration, we explicitly design the electric-circuit realization of a representative phase. 
Besides the immediate experimental interest, our work evidences that modulating gauge degrees of freedom may lead to unprecedented mechanisms and phenomena in non-Hermitian physics, which is a promising direction to explore.

\textit{\textcolor{blue}{Breaking symmetry restrictions by $\mathbb{Z}_{2}$
gauge fields.}}\textit{\textemdash{}}The modification of symmetry algebras with additional phase factors resembles the Aharonov-Bohm effect. Hence, to facilitate the minus sign in \eqref{eq:Switch_T}, it is sufficient to consider $\pi$-fluxes with gauge connections valued in $\mathbb{Z}_{2}=\{\pm 1\}$, i.e., each hopping amplitude takes phases $0$ or $\pi$ and therefore is a positive or negative real number. 

Then, for a plaquette with flux $\pi$, there are odd negative ones among the hopping amplitudes surrounding it. However, as illustrated in Figs.~\ref{fig:Fig1}(b) and (c), although the $\pi$-flux preserves $P=C$ or $M_z$, the gauge-connection configuration $A$ does not in general. $P$ transforms it to another configuration $A'$ that equally describes the $\pi$-flux configuration. Hence, the two gauge connection configurations $A$ and $A'$ are related by a gauge transformation $\mathsf{G}$, i.e., to restore the original $A$, a gauge transformation $\mathsf{G}$ should be imposed. Then, the parity operator that operates on the Hamiltonian should be the combination:
\begin{equation}
	\mathcal{P}=\mathsf{G}P
\end{equation}
Since $[T,\mathsf{G}]=[T,P]=0$, to realize \eqref{eq:Switch_T} we require $\mathcal{P}^2=-1$. This in turn requires the anticommutation relation $\{\mathsf{G},P\}=0$, noting that $P^2=\mathsf{G}^2=1$. It is noteworthy that $\mathsf{G}$ is just a diagonal matrix with diagonal entries $\pm 1$, where the matrix indexes are just the lattice sites. Thus, $\mathsf{G}$ can be visualized on the lattice by assigning each lattice site the corresponding sign (see Fig.~\ref{fig:Fig1}). Then, $\{\mathsf{G},P\}=0$ implies $P$ reverses the signs of all lattice sites. 

Particularly, let us consider the lattice blocks for $P=C$ and $M_z$, respectively, illustrated in Figs.~\ref{fig:Fig1}(b) and (c). For $P=C$ illustrated in Fig.~\ref{fig:Fig1}(b), we see that $C=\tau_1\otimes\sigma_1$ and $\mathsf{G}=\tau_3\otimes\sigma_0$, and clearly $C$ reverses signs of $\mathsf{G}$. Here, $\tau$ and $\sigma$ are two sets of the standard Pauli matrices operating on the row and column indexes, respectively. Hence, $\{C,\mathsf{G}\}=0$, and
\begin{equation}
	\mathcal{C}=\mathsf{G}C=i\tau_2\otimes\sigma_1
\end{equation}
with $\mathcal{C}^2=-1$.  For the mirror reflection, namely $P=M_z$, the operators are read off from Fig.~\ref{fig:Fig1}(c) as $\mathsf{G}=\tau_3\otimes\sigma_0$ and $M_z=\tau_1\otimes\sigma_0$. Accordingly, we have
\begin{equation}\label{eq:mirror}
	\mathcal{M}_z=i\tau_2\otimes\sigma_{0}
\end{equation}
with $\mathcal{M}_z^2=-1$.  

Then, the gauge-field enriched time-reversal symmetry is given by $\tilde{\mathcal{T}}=\mathcal{P}T$ with $\mathcal{P}=\mathcal{C}$ or $\mathcal{M}_z$. In momentum space, $T=U_TKI$ with $I$ the inversion of momenta and $K$ the complex conjugation. The non-Hermitian Hamiltonian $\mathcal{H}(\mathbf{k})$ in the ramified symmetry classes is constrained by $\tilde{\mathcal{T}}=\tilde{U}_TKI$ as
\begin{equation}
	\tilde{U}_T\mathcal{H}(\mathbf{k})^* \tilde{U}_T^\dagger=\mathcal{H}(-\mathbf{k})^\dagger. \label{eq:ant-sym}
\end{equation} 
Here, $\tilde{U}_T=\mathcal{P} U_T$. In addition to the ramified time reversal, we also consider the ramified sublattice symmetry $\mathcal{S}$, which exerts the following constraint on the Hamiltonian:
\begin{equation}
	\mathcal{S}\mathcal{H}^\dagger(\mathbf{k}) \mathcal{S}=-\mathcal{H}(\mathbf{k}). 
\end{equation}
Here, $\mathcal{S}$ is a unitary operator with $\mathcal{S}^2=1$. Hence, $\mathcal{S}$ is Hermitian unitary with $\mathcal{S}^\dagger=\mathcal{S}$.

\textit{\textcolor{blue}{1D spinful non-Hermitian topological phase 
		and $\mathbb{Z}_{2}$ skin effect
in spinless chains.}}\textit{\textemdash{}}Let us first consider
1D systems. According to the classification table of Fig.~\ref{fig:Fig1}(a), the emergence of any non-trivial topological phase is forbidden in 1D spinless systems with $T^2=+1$. Thus, it was previously concluded that the non-Hermitian skin effect, which manifests as the 
non-Hermitian bulk-boundary correspondence, is not allowed in these 
systems~\citep{yi2020nonhermitian,okuma2023nonhermitian}. On the 
contrary, non-trivial spinful topology is permitted in 1D. Here, 
we show that non-trivial topology in both of the two 1D spinful classes (AII$^\dagger$ 
and DIII$^\dagger$) in table~\ref{fig:Fig1}(a) can be constructed using the rectangle with $\pi$ flux of Fig.~\ref{fig:Fig1}(b). 

We focus on the topological phase in AII$^\dagger$ class, which possesses a spinful 
$\mathbb{Z}_2$ non-Hermitian skin effect, and leave the DIII$^\dagger$ class 
in the Supplemental Material (SM)~\cite{SuppInf}. We can build the 1D lattice model as shown in Fig.~\ref{fig:Fig2}(a). 
The momentum-space Hamiltonian reads, 
\begin{align}
	\mathcal{H}(k) & =J_{R}\tau_{0}\otimes\sigma_{1}-J_{I}\tau_{1}\otimes\sigma_{3}+t\cos k\tau_{0}\otimes \sigma_1\nonumber\\
	&+t\sin k \tau_{0}\otimes\sigma_{3} +\mu\tau_{0}\otimes\sigma_{2}+i\gamma\tau_{0}\otimes\sigma_{1},
\end{align}
where $J_R$ and $J_I$ are hoppings inside unit cells, $t$ denotes the hopping between unit cells, $\mu$ is a constant potential, and $i\gamma\tau_{0}\otimes\sigma_{1}$ is responsible for non-Hermiticity. We plot the 
energy spectra of the system under open boundary condition (OBC) and periodic
boundary condition (PBC) in Fig.~\ref{fig:Fig2}(c). 

Figure~\ref{fig:Fig2}(b) shows the wavefunction profile of the system 
under OBC. Evidently, a spinful $\mathbb{Z}_2$ 
non-Hermitian skin effect emerges in this spinless chain, indicating 
the breakdown of symmetry restrictions on 
this 1D spinless system. As previously explained, this is achieved by 
the gauge-field enriched time-reversal symmetry in Eq.~\eqref{eq:ant-sym} as
\begin{equation}
	\tilde{\mathcal{T}}=\mathcal{C}T= i\tau_2\otimes\sigma_1K I,\quad \tilde{\mathcal{T}}^2=-1, \label{eq:1Dtime}
\end{equation}
which leads
to a $\mathbb{Z}_{2}$ topological invariant $\nu(E)\in\{0,1\}$ in 1D~\citep{okuma2020topological},
\begin{align}
&(-1)^{\nu(E)}  =\sgn\Bigg\{\frac{\pf[(\mathcal{H}(\pi)-E)\tilde{U}_{T}]}{\pf[(\mathcal{H}(0)-E)\tilde{U}_{T}]}\nonumber \\
&\quad\times\exp\big(  -\frac{1}{2}\int_{k=0}^{k=\pi}d\log\det\text{\ensuremath{[(\mathcal{H}(k)-E})}\tilde{U}_{T}]\big)\Bigg\}.
\end{align}
Here, $\tilde{U}_{T}=i\tau_{2}\otimes\sigma_{1}$, $\pf$ denotes the Pfaffian of skew symmetric matrices~\cite{nakahara2018geometry}, and $E$
is a reference energy. We find the non-trivial $\nu(E)=1$ for $E$ inside the closed
curve of PBC spectrum. This topological invariant dictates the emergence of $\mathbb{Z}_{2}$
non-Hermitian skin effect localized at both ends of the OBC systems as shown in 
Fig.~\ref{fig:Fig2}(b).

The spinful $\mathbb{Z}_{2}$ skin effect can be understood by using the generalized
Brillouin zone (GBZ) \citep{WangPRL2018,yokomizo2019nonbloch,yang2020nonhermitian},
which is a generalization of the Bloch Hamiltonian by the substitution
of $e^{ik}\rightarrow\beta$ in $\mathcal{H}(k)$ to describe non-Hermitian
OBC systems. The non-Hermitian skin effect emerges when $|\beta|\neq|e^{ik}|=1$.
Note that skin modes with $|\beta|>1$ and those with $|\beta|<1$
are localized at opposite ends. Under the spinful time-reversal
symmetry of Eq.~(\ref{eq:1Dtime}), the skin mode
$|\beta,+\rangle$ is mapped to $|1/\beta,-\rangle$ in a Kramers
pair \citep{yi2020nonhermitian,kawabata2020nonbloch}. Here, $\pm$
denotes the two pseudo-spin states. The GBZ for symmetry-related pseudo-spin
bands are shown by red and blue curves in Fig.~\ref{fig:Fig2}(d),
which have inverse $\beta$ to each other. Thus, the spinful $\mathbb{Z}_{2}$
non-Hermitian skin effect is formed by the Kramers pairs of skin modes
localized at opposite ends.

\begin{figure*}[t]
	\includegraphics[width=\textwidth]{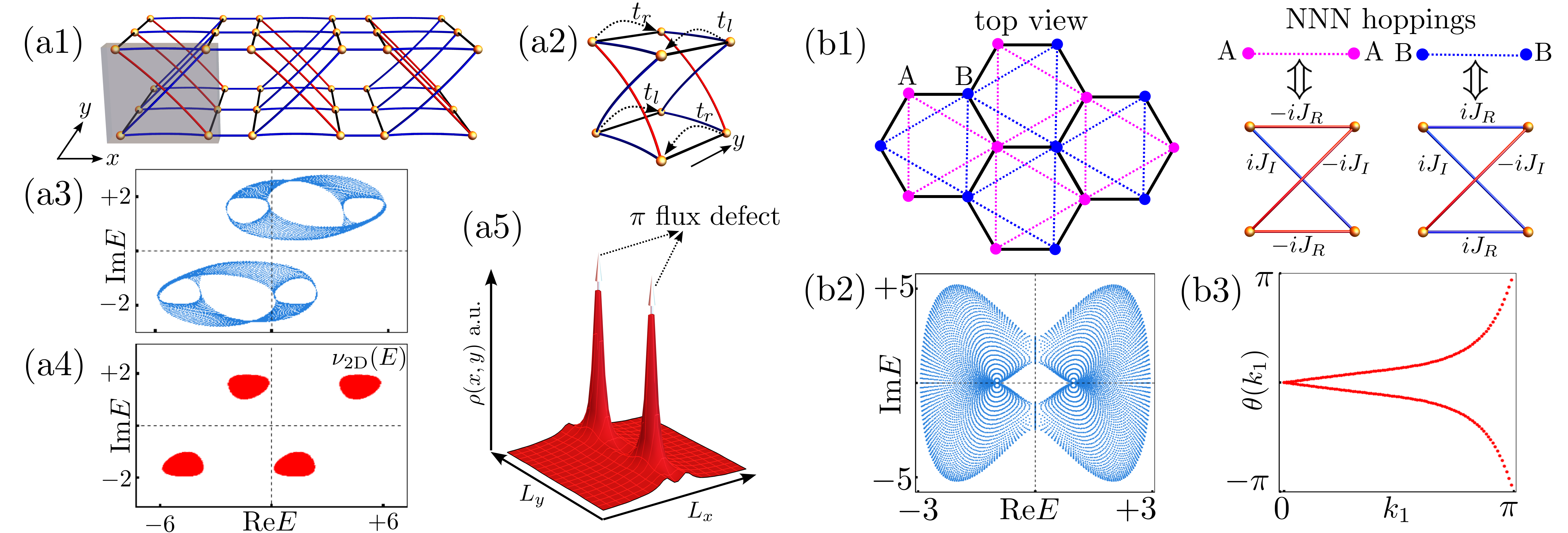}
	\caption{(a1) Schematic picture for the 2D lattice in AII$^{\dagger}$ class. The unit cell marked in gray. (a2) The non-reciprocal hoppings in the $y$ direction. (a3)
		The PBC spectrum, and (a4) the topological invariant of the system. 
		(a5) The non-Hermitian flux skin effect, which
		are localized at the cores of the $\pi$ flux defects. At the $(x,y)$
		position, the density is $\rho(x,y)=\sum_{\alpha,\text{\ensuremath{\beta}}}|\langle x,y,\alpha|\psi_{\beta}\rangle|^2$,
		where $\alpha$ runs over all internal degrees of unit cell and $\beta$
		all eigenstates. Full PBC is taken to avoid boundary effects. The parameters are
		$J_{R}=J_{I}=1.5,\gamma=t_{r}=2, t=t_{l}=1$. (b1) Schematic picture 
		for the bilayer hexagonal lattice. The two
		layers are connected by lattice blocks adjusted from Fig.~\ref{fig:Fig1}(c), as shown in the right. (b2) The PBC spectrum of the system, and (b3) the Wilson loop spectrum. The parameters are $t=J_{I}=1,J_{R}=0.1$.\label{fig:Fig3}}
\end{figure*}

\textit{\textcolor{blue}{2D non-Hermitian spinful topological phases and flux skin effect in spinless
lattices.}}\textit{\textemdash{}}We proceed to discuss the realization of 2D spinful topological
phases in spinless lattices. Among the three non-trivial 2D spinful classes in the lower table of Fig.~\ref{fig:Fig1}(a), we discuss two classes of AII$^\dagger$ and CII$^\dagger$ in the following, and leave DIII$^\dagger$ class 
in the SM~\citep{SuppInf}. Notably, for AII$^\dagger$ class in 2D, a novel type of 
non-Hermitian flux skin effect can emerge, which is different from the $\mathbb{Z}_2$ 
skin effect discussed above. Moreover, we propose the electric-circuit realization 
of the topological phase in this class in the SM~\citep{SuppInf}.

For all three spinful classes in 2D, we can use the twisted rectangle with $\pi$ flux in Fig.~\ref{fig:Fig1}(c) 
to construct their topological phases. In such a structure, the mirror symmetry is projectively represented as in Eq.~\eqref{eq:mirror}. Thus, the 
gauge-field enriched time-reversal symmetry is 
\begin{equation}\label{eq:2Dtime}
	\tilde{\mathcal{T}}=\mathcal{M}_zT= i\tau_2\otimes\sigma_0K I,\quad \tilde{\mathcal{T}}^2=-1, 
\end{equation}
which realizes the required symmetry algebra of Eq.~\eqref{eq:Switch_T}.

Our first 2D example is the spinful topological phase in class AII$^\dagger$ with  only 
time-reversal symmetry. It possesses the non-Hermitian flux skin effect as 
its characteristic feature~\cite{okuma2023nonhermitian,denner2022magnetic}. That is, when flux defects
are present, extensive numbers of skin modes will be localized at the flux cores.
As shown Fig.~\ref{fig:Fig3}(a1), this model is constructed by the block of Fig.~\ref{fig:Fig1}(c), which reads,
\begin{align}
	&\mathcal{H}_{\text{2D}}(k_{x},k_{y})  =\mathcal{H}(k_x) +\mathcal{H}(k_y),\nonumber\\
	&\mathcal{H}(k_y)= (t_{l}+t_{r})\cos k_{y}\tau_{0}\otimes\sigma_{0}+  i(t_{l}-t_{r})\sin k_{y}\tau_{3}\otimes\sigma_{0},\nonumber\\
	&\mathcal{H}(k_x)  =J_{R}\tau_{0}\otimes\sigma_{1}+J_{I}\tau_{2}\otimes\sigma_{2}+t\cos k_x\tau_{0}\otimes \sigma_1\nonumber \\
	 &\qquad\quad+t\sin k_x\tau_{0}\otimes \sigma_2 +i\gamma\tau_{0}\otimes\sigma_{1},\label{eq:2Dmodel}
\end{align}
Here, $\mathcal{H}(k_x)$ denotes the Hamiltonian in the $x$ direction, which is composed of the $J_R$ and $J_I$ terms within unit cells, the hopping term $t$, and the $\gamma$ term to induce non-Hermiticity. In the $y$ direction, $\mathcal{H}(k_y)$ consists of non-reciprocal hoppings as shown in Fig.~\ref{fig:Fig3}(a2).  The hopping amplitudes are $t_{r}$ towards right and $t_{l}$ ($\neq t_{r}$) towards left on the top layer, while they are exchanged on the bottom layer. 
The PBC energy spectrum is plotted in Fig.~\ref{fig:Fig3}(a3).

While $\tilde{\mathcal{T}}$ in Eq.~\eqref{eq:2Dtime} remains to be 
the time-reversal symmetry 
for this 2D system, the individual $\mathcal{M}_z$ and $T$ symmetries are broken by the non-reciprocal hoppings in the $y$ direction. 
Corresponding to $\tilde{\mathcal{T}}$, a $\mathbb{Z}_{2}$
invariant $\nu_{\text{\text{2D}}}(E)\in\{0,1\}$ is defined in Refs.
\citep{okuma2023nonhermitian,kawabata2019symmetry}, where $E$ is
a reference energy. We plot $\nu_{\text{\text{2D}}}(E)$ against $E$
in Fig.~\ref{fig:Fig3}(a4). The topological invariant takes non-trivial
values for $E$ inside the red region, indicating the 2D system is
topological.

Different from the 1D case, there is no $\mathbb{Z}_{2}$
skin effect under full OBC. Instead, a non-trivial magnetic flux response
serves as the charateristic feature of this spinful 2D topology~\citep{okuma2023nonhermitian,denner2022magnetic}.
By inserting a pair of $\pi$ flux defects, there will be $\mathcal{O}(L)$
skin modes from the total $\mathcal{O}(L^{2})$ modes localized
at the flux cores, where $L^2$ is the system size. In Fig.~\ref{fig:Fig3}(a5), the
2D system clearly shows such a non-Hermitian flux skin effect, 
confirming the non-trivial topology.
We provide the calculation details and the finite-size scaling of
the flux skin effect in the SM \citep{SuppInf}.

Our second example is a non-Hermitian generalized spinless Kane-Mele
model, which belongs to CII$^\dagger$ class in 2D. 
It is constructed based on the renowned
Kane-Mele model, where the spin-orbit coupling is replaced by
the spinless block of Fig.~\ref{fig:Fig1}(c) with adjustments~\cite{note}, as shown by the 
next-nearest-neighbour (NNN) hoppings in Fig.~\ref{fig:Fig3}(b1).
These blocks connect the top and bottom layers of the hexagonal lattice. 
The momentum-space Hamiltonian
reads,
\begin{align}
\mathcal{H}_{\text{KM}}(\mathbf{k}) & =t\chi_{1}(\mathbf{k})\tau_{0}\otimes\sigma_{1}+t\chi_{2}(\mathbf{k})\tau_{0}\otimes\sigma_{2}\nonumber \\
 & -2iJ_{R}\eta_{1}(\mathbf{k})\tau_{0}\otimes\sigma_{3}-2iJ_{I}\eta_{2}(\mathbf{k})\tau_{2}\otimes\sigma_{0},
\end{align}
where $t$ is the nearest-neighbour hopping amplitude, $\chi_{1}+i\chi_{2}=1+e^{ik_{1}}+e^{ik_{2}}$,
and $\eta_{1}+i\eta_{2}=e^{ik_{1}}+e^{ik_{2}}+e^{i(k_{2}-k_{1})}$
with $k_{1(2)}=\mathbf{k}\cdot\mathbf{b}_{1(2)}$. The primitive vectors
are $\mathbf{b}_{1(2)}=(3/2,\pm\sqrt{3}/2)$. We plot the PBC spetrum
in Fig.~\ref{fig:Fig3}(b2), which has a point gap with respect
to the zero energy.

Besides the time reversal symmetry of Eq.~\eqref{eq:2Dtime}, the system is invariant under the sublattice symmetry $\mathcal{S}=\tau_{0}\otimes\sigma_{3}$
as $\mathcal{S}\mathcal{H}_{\text{KM}}(\mathbf{k})^{\dagger}\mathcal{S}^{-1}=-\mathcal{H}_{\text{KM}}(\mathbf{k})$.
A 2D topological phase satisfying these symmetries is characterized by a $\mathbb{Z}_{2}$ Kane-Mele invariant \citep{kawabata2019symmetry}.
This invariant can be computed by the extended Hermitian Hamiltonian
$\mathcal{H}_{H}(\mathbf{k})=\antidiag\{\mathcal{H}_{\text{KM}}^{\dagger}(\mathbf{k}),\mathcal{H}_{\text{KM}}(\mathbf{k})\}$~\citep{SuppInf},
where $\antidiag$ denotes anti-diagonal matrice, due to the topological equivalence~\citep{Roy2017,gong2018,kawabata2019symmetry}. We then calculate the invariant using the Wilson loop spectrum~\citep{yu2011wilson}. A non-trivial (trivial)
topological invariant corresponds to odd (even) number of crossings
of the $\theta=\pi$ reference line. As shown in Fig.~\ref{fig:Fig3}(b3),
this spinful phase possesses a nontrivial $\mathbb{Z}_{2}$ invariant. 
More discussions on the topological edge states can be found in the SM~\citep{SuppInf}.

\textit{\textcolor{blue}{Summary and Discussions.}}\textit{\textemdash{} }In this work, we have presented mechanisms for realizing spinful ramified symmetry classes with the time-reversal symmetry by spinless systems, and systematically constructed experimentally realizable models for all topologically nontrivial cases in one and two dimensions. Our theory may be extended along the two directions, namely all non-Hermitian symmetry classes that contain either original or ramified time-reversal symmetry among the 38-fold symmetry classification, and non-Hermitian systems with other types of energy gaps~\cite{gapcondition}. This is due to the fact that the essence of our approach relies on the modification of symmetry algebras using gauge fields, which is independent of non-Hermitian ramification and complex-energy gaps. Our work shows that ubiquitously existing gauge structures may significantly enrich non-Hermitian physics at the fundamental level, and considering effects of gauge structures would be a promising direction to extend the current framework of non-Hermitian physics.

\vspace{0.5 cm}
\begin{acknowledgements} This work was supported by the Guangdong-Hong Kong Joint Laboratory of Quantum Matter, the NSFC/RGC JRS grant (RGC Grant No. N\_HKU774/21, NSFC Grant No. 12161160315), the CRF (Grant No. C6009-20G) and GRF (Grant No. 17310622) of Hong Kong, and the Basic Research Program of Jiangsu Province (Grant No. BK20211506). W. B. R. was supported by the RGC Postdoctoral Fellowship (Ref. No. PDFS2223-7S05). \end{acknowledgements}

\bibliography{Reference}

\begin{thebibliography}{54}
\bibitem [{com()}]{commutation}%
  \BibitemOpen
  \href@noop {} {}\bibinfo {note} {{The usual convention for the eight real
  Altland-Zirnbauer symmetry classes is to specify $T^2=\eta_T=\pm 1$ and
  $C^2=\eta_C=\pm 1$. Our convention is to specify $T^2=\eta_T$ and
  $TS=\alpha_{S}ST$ with $\alpha_S=\pm 1$, because sublattice symmetry $S$ is
  natural for non-superconducting systems. One may use $C=ST$ to translate the
  two conventions into each other }}\BibitemShut {NoStop}%
\bibitem [{not()}]{note}%
  \BibitemOpen
  \href@noop {} {}\bibinfo {note} {We note that the adjustments in the
  Fig.~\ref{fig:Fig3}(b1) right panel, i.e., NNN hoppings, do not break the
  $\tilde{\mathcal{T}}$ symmetry in Eq.~(\ref{eq:2Dtime}), and they turn the
  system to non-Hermitian. This is because for the NNN hoppings between A
  sublattices, we simply multiply a $-i$ coeffient to the original block of
  Fig.~\ref{fig:Fig1}(c); for the NNN hoppings between B sublattices, it is
  obtained by exchanging the two middle sticks(hoppings) between the top and
  bottom row of Fig.~\ref{fig:Fig1}(c), and then multiplied by $i$, which does
  not change the mirror-reflection symmetry $\mathcal{M}_z$.}\BibitemShut
  {Stop}%
\end{thebibliography}

\widetext
\clearpage


\section*{Supplementary material}

SUPPLEMENTAL MATERIAL FOR
"MAKING TOPOLOGICALLY TRIVIAL NON-HERMITIAN SYSTEMS NON-TRIVIAL VIA GAUGE FIELDS"

In this Supplemental Material, we construct the representative models for the rest non-trivial spinful symmetry classes in the classification table of Fig. 1(a) of the main text in Sec. I, propose the electric-circuit realization of the 2D model in class AII$^\dagger$ in Sec. II, propose an experimental realizable model with only on-site non-Hermiticity in Sec. III, provide the calculation details and the finite-size scaling of the non-Hermitian flux skin effect in Sec. IV, discuss the extended Hermitian Hamiltonian in Sec. V, use the extended Hermitian Hamiltonian to characterize the topology of the non-Hermitian generalized spinless Kane-Mele model in Sec. VI, and calculate the topological edge states of the non-Hermitian generalized spinless Kane-Mele model in Sec. VII.

## I. THE REST NON-TRIVIAL SPINFUL SYMMETRY CLASSES IN THE CLASSIFICATION TABLE

In the main text, we have constructed typical spinful topological phases in spinless systems for class AII$^\dagger$ in 1D and 2D, and CII$^\dagger$ in 2D. In the lower table of Fig. 1(a) of the main text, the rest two non-trivial spinful classes are class DIII$^\dagger$ in 1D and 2D, which have $\mathbb{Z}_2$ and $\mathbb{Z}$ type topological phases, respectively. Here, we construct the topological phases in these two classes, using the structures plotted in Fig. 1(b) and (c) in the main text.

### A. Class DIII$^\dagger$ in 1D

The topological phase in class DIII$^\dagger$ in 1D are of $\mathbb{Z}_2$ type. Using the $\pi$-flux rectangle in Fig. 1(b) in the main text, we can construct the model Hamiltonian as,

$$\mathcal{H}(k) = \underline{J_R \tau_0 \otimes \sigma_1 - J_I \tau_1 \otimes \sigma_3} + 2t \cos k \tau_0 \otimes \sigma_1 + 2it' \sin k \tau_2 \otimes \sigma_1 + i\gamma \tau_2 \otimes \sigma_3, \tag{S1}$$

where $t$ and $t'$ denote hoppings between the neighboring unit cells, and $\gamma$ is the loss and gain term within unit cells. Here, the underlined terms come from the hoppings inside the unit cell of the $\pi$-flux rectangle of Fig. 1(b) in the main text. The energy spectrum is plotted in Fig. S1 (a), which possesses a point gap with respect to the zero energy.

This system respects the symmetries of

$$\tilde{U}_T \mathcal{H}(k)^t \tilde{U}_T^\dagger = \mathcal{H}(-k), \quad \tilde{U}_T = i\tau_2 \otimes \sigma_1, \tag{S2}$$

$$\mathcal{S} \mathcal{H}(k)^\dagger \mathcal{S}^{-1} = -\mathcal{H}(k), \quad \mathcal{S} = \tau_1 \otimes \sigma_2. \tag{S3}$$

The two symmetries satisfy

$$\tilde{\mathcal{T}}^2 = -1, \quad \{\tilde{\mathcal{T}}, \mathcal{S}\} = 0, \tag{S4}$$

where $\tilde{\mathcal{T}} = \tilde{U}_T K I$. Thus, the system belongs to class DIII$^\dagger$ in 1D. For point gapped systems in this class, we can use the topological invariant $\nu \in \{0, 1\}$ of [1]

$$(-1)^\nu = \text{sgn}\left(\frac{\text{Pf}[\mathcal{H}(\pi)\tilde{U}_T]}{\text{Pf}[\mathcal{H}(0)\tilde{U}_T]}\right) \tag{S5}$$

to capture the $\mathbb{Z}_2$ topology. For the model Hamiltonian in Eq. (S1), it can be derived that

$$(-1)^\nu = \text{sgn}\left(1 - \frac{8 J_R t}{J_I^2 + (J_R + 2t)^2 - \gamma^2}\right). \tag{S6}$$

For the chosen parameters in Fig. S1, this topological invariant is $\nu = 1$, indicating a non-trivial topological phase.

### B. Class DIII$^\dagger$ in 2D

We procceed to construct the spinful model Hamiltonian for 2D systems in class DIII$^\dagger$, which is characterized by a $\mathbb{Z}$ type topological invariant. The momentum-space Hamiltonian reads,

$$\mathcal{H}(k_x, k_y) = i(\underline{J_R \tau_0 \otimes \sigma_1 + J_I \tau_2 \otimes \sigma_2}) + t \sin k_x \tau_1 \otimes \sigma_3 + t \sin k_y \tau_1 \otimes \sigma_1 - it(\cos k_x + \cos k_y) \tau_2 \otimes \sigma_2, \tag{S7}$$

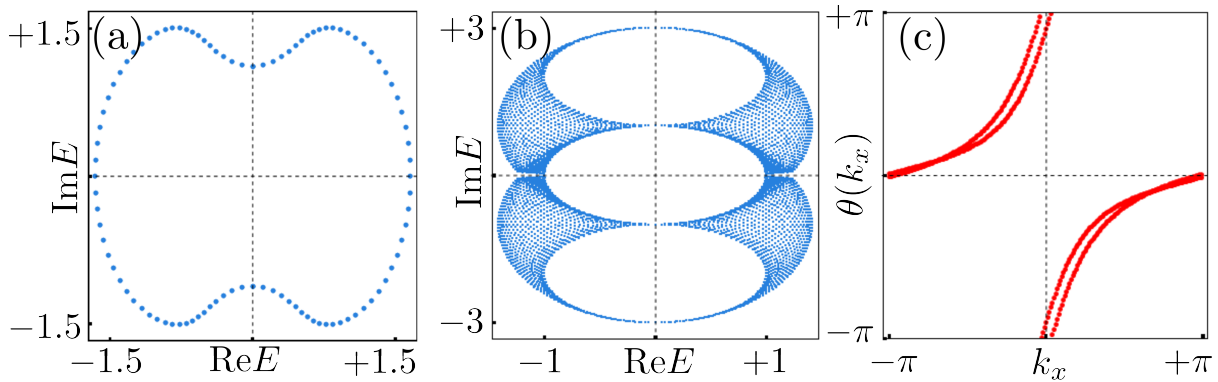

Figure S1. (a) The energy spectrum under PBC for the model Hamiltonian Eq. (S1). The parameters are $t = t' = 0.5, J_R = J_I = 1, \gamma = 1.4$. (b) The energy spectrum for the model Hamiltonian Eq. (S7). (C) The Wilson loop spectrum of $i\mathcal{H}(k_x, k_y)\mathcal{S}$ for the two valence bands. The parameters are $J_R = 0.1, J_I = 1, t = 1$.

where $\sin k_{x(y)}$ and $\cos k_{x(y)}$ terms come from the hoppings between unit cells (underlined, twisted rectangle in Fig. 1(c) in the main text) in $x(y)$ directions. Note that same as the non-Hermitian generalized spinless Kane-Mele model in the main text, the hopping amplitudes within each unit cell are multiplied by a coefficient of $i$, which does not break the underlying symmetry. The PBC energy spectrum is plotted in in Fig. S1(b), which possess a point gap about the zero energy.

This system respects the symmetries of

$$\tilde{U}_T \mathcal{H}(k_x, k_y)^t \tilde{U}_T^\dagger = \mathcal{H}(-k_x, -k_y), \quad \tilde{U}_T = i\tau_2 \otimes \sigma_0, \tag{S8}$$

$$\mathcal{S}\mathcal{H}(k_x, k_y)^\dagger \mathcal{S}^{-1} = -\mathcal{H}(k_x, k_y), \quad \mathcal{S} = \tau_2 \otimes \sigma_0, \tag{S9}$$

where $\tilde{\mathcal{T}} = \tilde{U}_T K I$. The two symmetries also satisfy

$$\tilde{\mathcal{T}}^2 = -1, \quad \{\tilde{\mathcal{T}}, \mathcal{S}\} = 0. \tag{S10}$$

Thus, the system belongs to class DIII$^\dagger$ in 2D. The point-gapped 2D non-Hermitian system in class DIII$^\dagger$ is of $\mathbb{Z}$ classification, and is characterized by the Chern number [1].

To obtain the topological invariant, it suffices to compute the Chern number of the corresponding Hermitian Hamiltonian $i\mathcal{H}(k_x, k_y)\mathcal{S}$, which satisfies $(i\mathcal{H}(k_x, k_y)\mathcal{S})^\dagger = i\mathcal{H}(k_x, k_y)\mathcal{S}$ [1]. In Sec. V, we will show that this is because $i\mathcal{H}(k_x, k_y)\mathcal{S}$ captures the topology of the extended Hermitian Hamiltonian of anti-diag$\{\mathcal{H}(k_x, k_y)^\dagger, \mathcal{H}(k_x, k_y)\}$, which is topolgically equivalent to the non-Hermitian Hamiltonian $\mathcal{H}(k_x, k_y)$.

As shown in Fig. S1(c), we compute the Wilson loop spectrum of $i\mathcal{H}(k_x, k_y)\mathcal{S}$ for the two valance bands. The Chern number can be obtained by the winding number of the Wannier center $\theta(k_x)$ of the two bands. Clearly, the total Chern number is 2, as both valence bands wind once in one period of $k_x \in (-\pi, \pi]$.

## II. EXPERIMENTAL REALIZATION USING ELECTRIC CIRCUITS

Before going to concrete models, here we briefly illustrate the symmetry operations involved in the development of our theory. As shown in Fig. S2, for both 1D and 2D systems, $P$ can be the mirror reflection $M_z$ inversing the $z$ axis, presuming that the system is placed on the $x-y$ plane. For a 1D system, we may choose $P$ as the twofold rotation $C$ through the direction of the 1D system. We can construct models satisfying these symmetries using artificial crystals.

Now let us discuss the experimental realization of concrete models. The proposed lattice models in the main text can be engineered in various artificial systems. Here, we consider the realization using electric-circuit arrays. In particular, we propose the realization of the twisted $\pi$-flux rectangle in Fig. 1(c) in the main text, and apply it to construct the 2D models in class AII$^\dagger$, i.e., Eq. (11) in the main text.

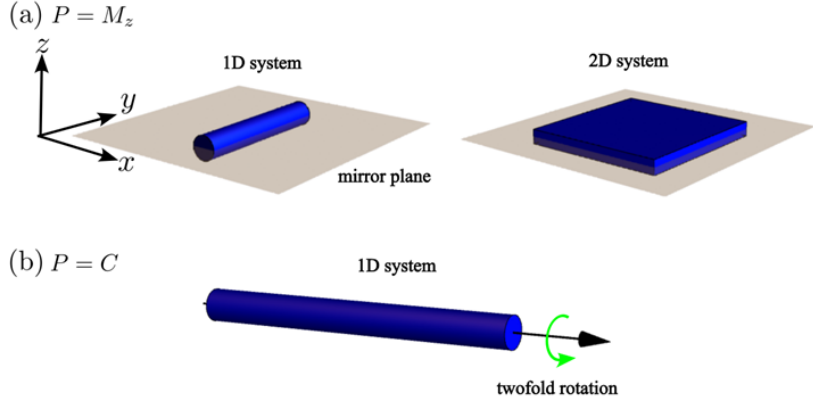

Figure S2. (a) For both 1D and 2D systems, $P$ could be mirror reflection $M_z$. (b) For 1D system, $P$ could also be the twofold rotation $C$ through the direction of the 1D system

Specifically, the targeted model Hamiltonian is

$$\mathcal{H}_{2D}(k_x, k_y) = \mathcal{H}(k_x) + \mathcal{H}(k_y),$$
$$\mathcal{H}(k_x) = \underline{J_R \tau_0 \otimes \sigma_1 + J_I \tau_2 \otimes \sigma_2} + t \cos k_x \tau_0 \otimes \sigma_1$$
$$+ t \sin k_x \tau_0 \otimes \sigma_2 + i\gamma \tau_0 \otimes \sigma_1,$$
$$\mathcal{H}(k_y) = (t_l + t_r) \cos k_y \tau_0 \otimes \sigma_0 + i(t_l - t_r) \sin k_y \tau_3 \otimes \sigma_0, \tag{S11}$$

where the underlined part comes from the twisted $\pi$-flux rectangle in Fig. 1(c) in the main text. We first construct the model Hamiltonian for the twisted $\pi$-flux rectangle, then the Hamiltonian $\mathcal{H}(k_x)$ in the $x$ direction, and then the $y$ direction Hamiltonian of $\mathcal{H}(k_y)$.

For electric circuits, the Kirchhoff rule in the frequency domain is,

$$I(\omega) = J(\omega) V(\omega), \tag{S12}$$

where $I$ denotes the current and $V$ the voltage measured against ground at a node, respectively. Here, $J$ is the admittance, and $\omega$ is the AC driving frequency of the system. Specifically, consider two nodes denoted by $a$ and $b$. They could be connected by a capacitor $C$, an inductor $L$, or a resistor $R$, and the Kirchhoff rules for them are

$$I_a = i\omega C (V_a - V_b),$$
$$I_a = \frac{1}{i\omega L}(V_a - V_b),$$
$$I_a = \frac{1}{R}(V_a - V_b). \tag{S13}$$

### A. The twisted rectangle with $\pi$ flux

To realize the twisted $\pi$-flux rectangle structure in Fig. 1(c) in the main text, we propose the structure as shown in Fig. S3(a), which is made up of capacitors ($C$) and inductors ($L$). Then from Eqs. (S12) and (S13), the curent-voltage relation reads

$$\begin{pmatrix} I_A \\ I_B \\ I_C \\ I_D \end{pmatrix} = \begin{pmatrix} i\omega C + \frac{1}{i\omega L_1} & -i\omega C & 0 & -\frac{1}{i\omega L_1} \\ -i\omega C & i\omega(C + C_1) & -i\omega C_1 & 0 \\ 0 & -i\omega C_1 & i\omega(C + C_1) & -i\omega C \\ -\frac{1}{i\omega L_1} & 0 & -i\omega C & i\omega C + \frac{1}{i\omega L_1} \end{pmatrix} \begin{pmatrix} V_A \\ V_B \\ V_C \\ V_D \end{pmatrix}. \tag{S14}$$

Here, we use $A, B, C, D$ to denotes the four nodes [yellow points in Fig. S3(a)]. The corresponding admittance matrix is

$$J_0(\omega) = i\omega D_0(\omega) - i\omega H_0(\omega), \tag{S15}$$

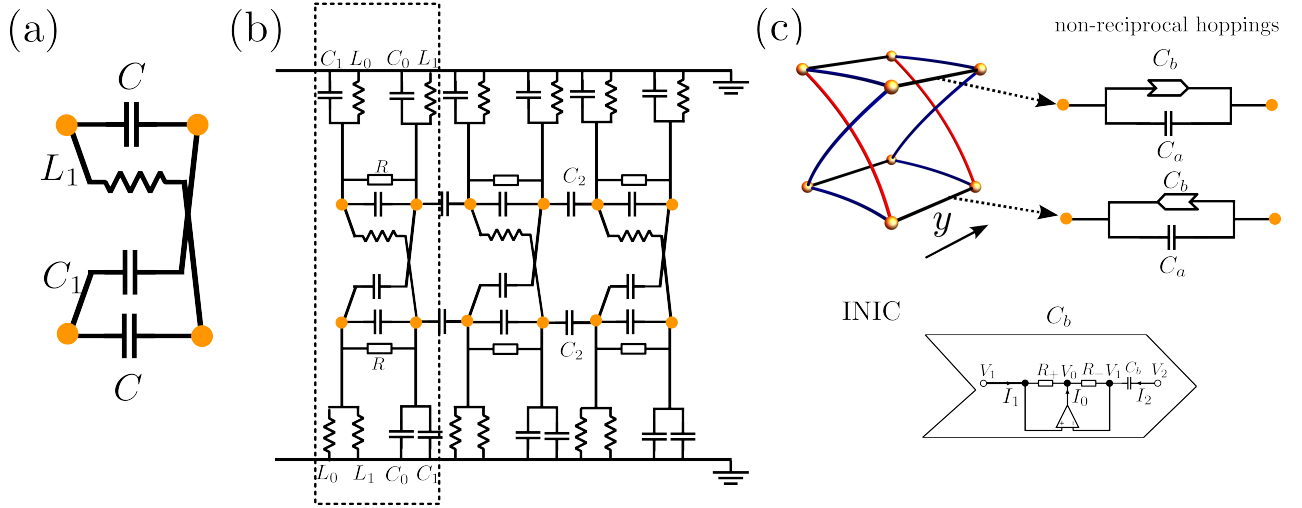

Figure S3. (a) The electric-circuit realization of the twisted rectangle with $\pi$ flux for Fig. 1(c) in the main text. (b) The $x$-directional Hamiltonian component $\mathcal{H}(k_x)$ in the 2D model of Class AII$^\dagger$ in Eq. (S11) can be realized by electric-circuit array. The unit cell is marked by the dashed square. (c) The non-reciprocal hoppings [$\mathcal{H}(k_y)$] of the 2D model in Class AII$^\dagger$ in the $y$ direction can be constructed by using the INIC components. Lower panel: the details of the INIC.

where $D(\omega)$ and $H(\omega)$ are,

$$D_0(\omega) = \begin{pmatrix} C - \frac{1}{\omega^2 L_1} & 0 & 0 & 0 \\ 0 & C + C_1 & 0 & 0 \\ 0 & 0 & C + C_1 & 0 \\ 0 & 0 & 0 & C - \frac{1}{\omega^2 L_1} \end{pmatrix}, \quad H_0(\omega) = \begin{pmatrix} 0 & C & 0 & -\frac{1}{\omega^2 L_1} \\ C & 0 & C_1 & 0 \\ 0 & C_1 & 0 & C \\ -\frac{1}{\omega^2 L_1} & 0 & C & 0 \end{pmatrix}. \tag{S16}$$

In $H(\omega)$, we can set

$$\omega_0 = 1/\sqrt{L_1 C_1}, \tag{S17}$$

i.e., $\frac{1}{\omega_0^2 L_1} = C_1$, and denote

$$J_R = C, \quad J_I = C_1. \tag{S18}$$

Then, we arrive at

$$H_0(\omega_0) = \begin{pmatrix} 0 & J_R & 0 & -J_I \\ J_R & 0 & J_I & 0 \\ 0 & J_I & 0 & J_R \\ -J_I & 0 & J_R & 0 \end{pmatrix} = J_R \tau_0 \otimes \sigma_1 + J_I \tau_2 \otimes \sigma_2. \tag{S19}$$

Clearly, $H_0(\omega_0)$ realizes the targeted Hamiltonian $H$ for the twisted $\pi$-flux rectangle. As for the the onsite term $D_0(\omega_0)$, it can be neglected by engineering the grounding elements. In the following two sections, we disscuss the realization of the $x$ and $y$ direction components in the 2D model of Eq. (S11).

### B. The Hamiltonian component $\mathcal{H}(k_x)$ in the $x$ direction

Here, we propose the electric-circuit realization of $\mathcal{H}(k_x)$ component in Eq. (S11). In the Hamiltonian, $t$ is the hopping between unit cells, $\mu$ is the onsite potential, and $\gamma$ the term makes the system non-Hermitian.

As shown by Fig. S3(b), the model Hamiltonian of $\mathcal{H}(k_x)$ can be constructed using the unit cell made up by the electric circuit in (a). Besides the unit cell $J_0(\omega)$, there are the hopping terms between unit cells ($C_2$) and hopping

terms within unit cells ($R$), which correspond to the current-voltage relation of

$$\begin{pmatrix} I_A(x) \\ I_B(x) \\ I_C(x) \\ I_D(x) \end{pmatrix} = i\omega C_2 \begin{pmatrix} V_A(x) - V_B(x-1) \\ V_B(x) - V_A(x+1) \\ V_C(x) - V_D(x-1) \\ V_D(x) - V_C(x+1) \end{pmatrix}, \quad \begin{pmatrix} I_A(x) \\ I_B(x) \\ I_C(x) \\ I_D(x) \end{pmatrix} = \frac{1}{R} \begin{pmatrix} V_A(x) - V_B(x) \\ V_B(x) - V_A(x) \\ V_C(x) - V_D(x) \\ V_D(x) - V_C(x) \end{pmatrix}, \quad \text{(S20)}$$

where $x$ is the position of the node. The grounding elements are

$$\begin{pmatrix} I_A(x) \\ I_B(x) \\ I_C(x) \\ I_D(x) \end{pmatrix} = \begin{pmatrix} i\omega C_1 V_A(x) + \frac{1}{i\omega L_0} V_A(x) \\ i\omega C_0 V_B(x) + \frac{1}{i\omega L_1} V_B(x) \\ \frac{1}{i\omega L_0} V_C(x) + \frac{1}{i\omega L_1} V_C(x) \\ i\omega C_0 V_D(x) + i\omega C_1 V_D(x) \end{pmatrix}. \quad \text{(S21)}$$

After Fourier transformation, the momentum-space admittance model becomes

$$J(\omega) = i\omega D(\omega) - i\omega H(\omega), \quad \text{(S22)}$$

where

$$D(\omega) = D_0(\omega) + (C_2 + \frac{1}{i\omega R})\mathbb{1} + \text{diag}\{C_1, -C_1, -C_1, C_1\},$$
$$H(\omega) = H_0(\omega) + C_2 \tau_0 \otimes \begin{pmatrix} 0 & e^{-ik} \\ e^{ik} & 0 \end{pmatrix} - \frac{i}{\omega R} \tau_0 \otimes \sigma_1 + C_0 \tau_0 \otimes \sigma_3. \quad \text{(S23)}$$

Here "diag" denotes the diagonal matrix. We can set the AC frequence $\omega = \omega_0$ to make $L_0, C_0$ and $L_1 C_1$ satisfy the following conditions,

$$1/\sqrt{L_0 C_0} = 1/\sqrt{L_1 C_1} \equiv \omega_0. \quad \text{(S24)}$$

Thus, by denoting

$$C_2 = t, \quad -\frac{1}{\omega R} = \gamma, \quad C_0 = \mu, \quad \text{(S25)}$$

we have arrived at the desired Hamitonian component $\mathcal{H}(k_x)$ in Eq. (S11) from $H(\omega)$, by setting $\mu = 0$, i.e., neglecting the $C_0$ and $L_0$ components. Note that after the above treatment, $D(\omega) = (C + C_2 + \frac{1}{i\omega R})\mathbb{1}$ is proportional to the identity matrix, and thus, does not affect the topological properties of topological phases in this class.

### C. The Hamiltonian component $\mathcal{H}(k_y)$ in the $y$ direction

Here, we propose the electric-circuit realization of $\mathcal{H}(k_y)$ in Eq. (S11). The hopping terms in the $y$ direction are non-reciprocal ($t_l \neq t_r$), and they can be realized by INIC (negative impedance converter with current inversion) [2]. The current-voltage relation for such a device reads,

$$\begin{pmatrix} I_1 \\ I_2 \end{pmatrix} = i\omega C \begin{pmatrix} -1 & 1 \\ -1 & 1 \end{pmatrix} \begin{pmatrix} V_1 \\ V_2 \end{pmatrix}, \quad \text{(S26)}$$

where $I_1$, $V_1$ and $I_2$, $V_2$ are the current and voltage at the two ends, as shown by Fig. S3(c) lower panel. The mechanism of such a device can be found in Ref. [2]. Its compacity is $-C$ in the direction along the arrow, and $C$ in the direction against the arrow.

In the $y$ direction, for the hoppings between unit cells in the top layer, as shown by Fig. S3(c), the corresponding current-voltage relation are

$$\begin{pmatrix} I_A(y) \\ I_B(y) \end{pmatrix} = -i\omega C_b \begin{pmatrix} [V_A(y) - V_A(y+1)] - [V_A(y) - V_A(y-1)] \\ [V_B(y) - V_B(y+1)] - [V_B(y) - V_B(y-1)] \end{pmatrix} = -i\omega C_b \begin{pmatrix} -V_A(y+1) + V_A(y-1) \\ -V_B(y+1) + V_B(y-1) \end{pmatrix}, \quad \text{(S27)}$$

and in the bottom layer, they are

$$\begin{pmatrix} I_C(y) \\ I_D(y) \end{pmatrix} = i\omega C_b \begin{pmatrix} [V_C(y) - V_C(y+1)] - [V_C(y) - V_C(y-1)] \\ [V_D(y) - V_D(y+1)] - [V_D(y) - V_D(y-1)] \end{pmatrix} = i\omega C_b \begin{pmatrix} -V_C(y+1) + V_C(y-1) \\ -V_D(y+1) + V_D(y-1) \end{pmatrix}. \quad \text{(S28)}$$

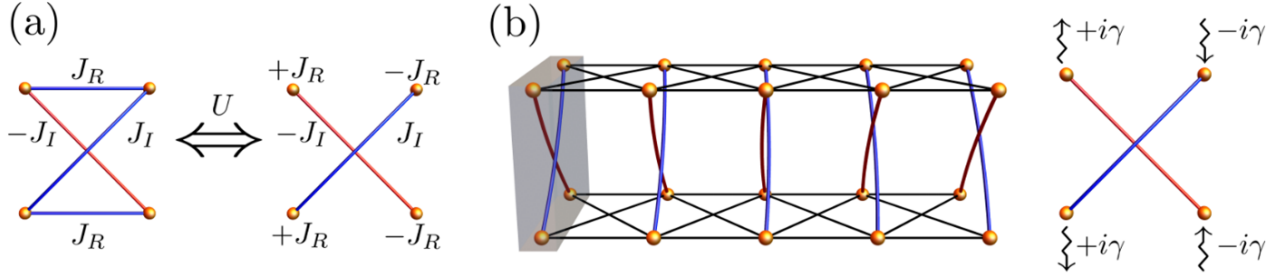

Figure S4. A non-Hermitian spinful model realized by a 1D spinless chain, which contains only positive/negative couplings and onsite non-Hermiticity. (a) The two blocks are equivalent to each other up to a unitary transformation. (b) The sketch for the tight-binding model. Non-Hermiticity is induced by only the onsite term.

Together with the normal capacitor $C_a$, the current-voltage relation for the hoppings in the $y$ direction are

$$\begin{pmatrix} I_A(y) \\ I_B(y) \\ I_C(y) \\ I_D(y) \end{pmatrix} = i\omega C_a \begin{pmatrix} 2V_A(y) \\ 2V_B(y) \\ 2V_C(y) \\ 2V_D(y) \end{pmatrix} - i\omega \begin{pmatrix} (C_a - C_b)V_A(y+1) + (C_a + C_b)V_A(y-1) \\ (C_a - C_b)V_B(y+1) + (C_a + C_b)V_B(y-1) \\ (C_a + C_b)V_C(y+1) + (C_a - C_b)V_C(y-1) \\ (C_a + C_b)V_D(y+1) + (C_a - C_b)V_D(y-1) \end{pmatrix}. \tag{S29}$$

After Fourier transformation, the momentum-space admittance model becomes

$$J_y(\omega) = i\omega D_y(\omega) - i\omega H_y(\omega), \tag{S30}$$

where $D_y(\omega) = 2C_a \mathbb{1}$, and

$$H_y(\omega) = \text{diag}\{t_l e^{ik_y} + t_r e^{-ik_y}, t_l e^{ik_y} + t_r e^{-ik_y}, t_r e^{ik_y} + t_l e^{-ik_y}, t_r e^{ik_y} + t_l e^{-ik_y}\}, \tag{S31}$$

where $t_l = C_a - C_b$, and $t_r = C_a + C_b$. We can see that $H_y(\omega)$ term realizes the terms containing $k_y$ in the 2D model of Eq. (S11), i.e., $H_y(\omega) = (t_l + t_r)\cos k_y \tau_0 \otimes \sigma_0 + i(t_l - t_r)\sin k_y \tau_3 \otimes \sigma_0$. In combination with the realization of the $\mathcal{H}(k_x)$ in the previous section, the 2D model $\mathcal{H}_{2D}(k_x, k_y)$ in Class AII$^\dagger$ can be experimentally realized.

### III. EXPERIMENTAL REALIZABLE MODEL WITH ONLY ON-SITE NON-HERMITICITY

In the main text, the proposed models consist of non-reciprocal couplings (or non-Hermiticity in couplings). With recent experimental advances, such couplings can be realized in different experimental platforms, such as electric circuits [2], photonic crystals [3], and acoustic crystals [4]. Hence, the concrete models we propose in this work are realizable using these techniques.

Nonetheless, in this section, we propose a concrete model that contains only on-site non-Hermiticity, which can make our work more accessible to experiments. Let us start with the twisted rectangle with $\pi$ flux [i.e., Fig. 1(c) in the main text], which we replot at left side of Fig. S4(a) below for convenience.

The model Hamiltonian for the twisted rectangle reads,

$$\mathcal{H} = J_R \tau_0 \otimes \sigma_1 + J_I \tau_2 \otimes \sigma_2. \tag{S32}$$

We can perform a unitary transformation $U$ on this Hamiltonian as

$$\mathcal{H}' = U\mathcal{H}U^{-1} = J_R \tau_0 \otimes \sigma_3 + J_I \tau_2 \otimes \sigma_2, \quad U = \exp\left(i\frac{\pi}{4}\tau_0 \otimes \sigma_2\right). \tag{S33}$$

The sketch for the model $\mathcal{H}'$ is plotted at the right side of Fig. S4(a), which is equivalent to the original $\mathcal{H}$ up to a unitary transformation. We note that $\mathcal{H}'$ is also invariant under the spinful time-reversal symmetry $\mathcal{T}' = U\mathcal{T}U^{-1} = i\tau_2 \otimes \sigma_0 KI$, which satisfies $\mathcal{T}'^2 = -1$.

Using the block of $\mathcal{H}'$ as the unit cell, we can construct the targeted non-Hermitian spinful model in class AII$^\dagger$ in 1D, as shown in Fig. S4(b). The momentum-space model Hamiltonian reads,

$$\mathcal{H}'(k) = J_R \tau_0 \otimes \sigma_3 + J_I \tau_2 \otimes \sigma_2 + 2t_1 \cos k \tau_0 \otimes \sigma_3 + 2t_2 \sin k \tau_0 \otimes \sigma_2 + i\gamma \tau_0 \otimes \sigma_3. \tag{S34}$$

There are three parts of this model Hamiltonian, i.e., intracell terms $(J_R, J_I)$, intercell terms $(t_1, t_2)$, and onsite non-Hermiticity $(i\gamma)$. The coupling terms (i.e., $J_I, t_1, t_2$) take only positive/negative values, and the onsite non-Hermiticity is introduced by including the non-Hermitian $i\gamma$ term to the original onsite $J_R$ term. In this way, the strict restrictions of only containing positive/negative couplings and onsite non-Hermiticity are fulfilled.

To characterize the topological phase, we can use the topological invariant $\nu(E)$ of Eq. (9) in the main text. We find a non-trivial topological invariant of

$$\nu(E) = 1, \tag{S35}$$

for the parameters $J_R = J_I = 1.5, t_1 = t_2 = 0.5, \gamma = 2$ at the reference energy $E = 2 + 1.5\,i$, indicating a non-trivial topological phase. Similar results of the spinful $\mathbb{Z}_2$ non-Hermitian skin effect as Fig. 2(b) in the main text can also be observed for these parameters.

## IV. 2D NON-HERMITIAN FLUX SKIN EFFECT IN CLASS AII$^\dagger$

In this section, we provide more calculation details and the finite-size scaling of the non-Hermitian flux skin effect of the 2D model of class AII$^\dagger$. As discussed in the previous section, the corresponding model of Eq. (S11) can be realized in artificial systems, such as the electric-circuit array.

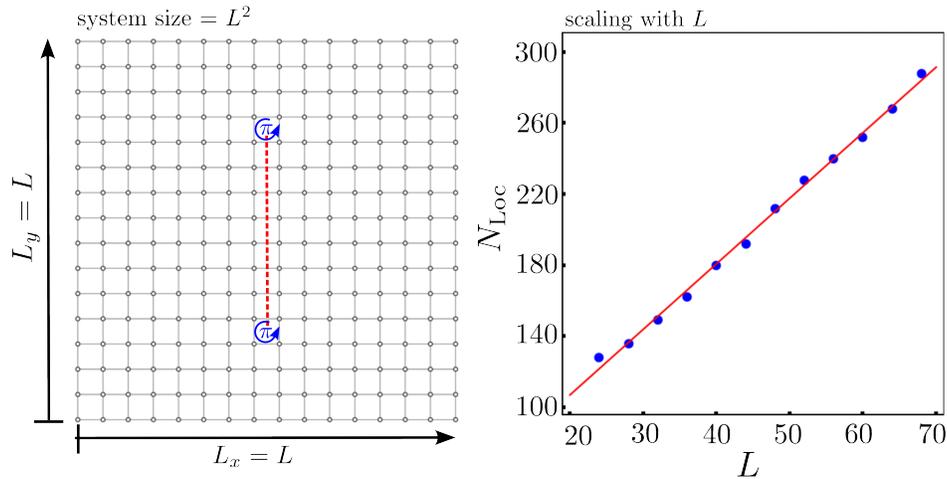

Figure S5. Left: Schematic picture for inserting a pair of $\pi$ flux defects on the 2D lattice. The $\pi$ flux defects are highlighted in blue. They are positioned at $(\frac{L_x}{2}, \frac{L_y}{4})$ and $(\frac{L_x}{2}, \frac{3L_y}{4})$. Right: The number of localized modes ($N_{\rm LOC}$) larger than 60% around the positions of the two $\pi$ flux defects. $N_{\rm LOC}$ scales with $L$ instead of the system size $L^2$. The fit (red line) of the numerical data (blue dots) is $N_{\rm LOC} = 33.1107 + 3.69143L$. The parameters are the same as in the main text, except that $L = 40$ for Fig. 3(a5) in the main text.

As shown in Fig. S5 left panel, on the 2D lattice, a pair of $\pi$ flux defects are inserted at the positions $(\frac{L_x}{2}, \frac{L_y}{4})$ and $(\frac{L_x}{2}, \frac{3L_y}{4})$, and the total system size is $L_x = L_y = L$. These $\pi$ fluxes are introduced by multiplying a phase factor of $e^{i\pi} = -1$ to the hoppings between $x = L_x/2$ and $x = L_x/2 + 1$ for $L_y/4 \leq y \leq 3L_y/4$, as highlighted by the red dashed line in Fig. S5 left panel. Note that we have taken PBC in both $x$ and $y$ directions to avoid boundary effects.

To observe the non-Hermitian flux skin effect, we compute the density at the position $(x, y)$ as

$$\rho(x, y) = \sum_{\alpha, \beta} |\langle x, y, \alpha | \psi_\beta \rangle|^2, \tag{S36}$$

where $\alpha$ runs over all sublattice degrees of freedom and $\beta$ runs over all eigenstates of the Hamitlonian with the $\pi$ flux defects. Note that all the eigenstates $|\psi_\beta\rangle$ are normalized before calculation. In the main text, we plot $\rho(x, y)$ against $(x, y)$ to show the non-Hermitian flux skin effect.

For the 2D system of class AII$^\dagger$, there are $\mathcal{O}(L)$ skin modes from the total $\mathcal{O}(L^2)$ modes, which are localized at the flux cores under full PBC. To confirm this, we can observe the scaling behavior by increasing the system size.

Consider the regions ($\mathcal{R}$) around the flux cores, which are

$$\mathcal{R}: \quad \begin{array}{ll} \text{flux core at } (\frac{L_x}{2}, \frac{L_y}{4}): & |x - \frac{L_x}{2}| \leq 4, \quad \text{and } |x - \frac{L_y}{4}| \leq 4, \\ \text{flux core at } (\frac{L_x}{2}, \frac{3L_y}{4}): & |x - \frac{L_x}{2}| \leq 4, \quad \text{and } |x - \frac{3L_y}{4}| \leq 4. \end{array} \tag{S37}$$

We denote these regions as $\mathcal{R}$, and sum up the weight of a eigenfunction in these regions. If the summation of the weight of a eigenfunction $\psi_\beta$ fulfills the criterion of

$$\sum_{(x,y)\in\mathcal{R}} \sum_\alpha |\langle x, y, \alpha | \psi_\beta \rangle|^2 \geq 60\%, \tag{S38}$$

then we count the number of localized modes ($N_{\text{LOC}}$) at the flux cores. As shown in Fig. S5 right panel, $N_{\text{LOC}}$ scales with $L$ instead of the system size $L^2$, showing that there are $\mathcal{O}(L)$ non-Hermitain flux skin modes from the total $\mathcal{O}(L^2)$ modes in the system.

## V. EXTENDED HERMITIAN HAMILTONIAN

In this section, we discuss the extended Hermitian Hamiltonian, constructed by the point-gapped non-Hermitian Hamiltonian. Let us consider a general point-gapped system $H_{\text{nh}}$ with non-Hermitian ramified chiral symmetry

$$\mathcal{S} H_{\text{nh}}^\dagger \mathcal{S} = -H_{\text{nh}}, \quad \mathcal{S}^2 = +1. \tag{S39}$$

We can take an appropriate basis, so that the Hamiltonian and the chiral symmetry operator take the form

$$\mathcal{H}_{\text{nh}} = \begin{pmatrix} h_a & h_b \\ h_c & h_d \end{pmatrix}, \quad \mathcal{S} = \begin{pmatrix} \mathbb{1} & \\ & -\mathbb{1} \end{pmatrix}, \tag{S40}$$

where $\mathbb{1}$ is the identity matrix with the dimension equal to half of $H_{\text{nh}}$. Due to the chiral symmetry in Eq. (S39), it is required that

$$h_a^\dagger = -h_a, \quad h_b^\dagger = h_c, \quad h_c^\dagger = h_b, \quad h_d^\dagger = -h_d. \tag{S41}$$

The point gap Hamiltonian $H_{\text{nh}}$ is topological equivalent to the extended Hermitian Hamiltonian of

$$\mathcal{H}_H = \begin{pmatrix} & \mathcal{H}_{\text{nh}} \\ \mathcal{H}_{\text{nh}}^\dagger & \end{pmatrix}, \quad \mathcal{S}_H = \begin{pmatrix} & \mathcal{S} \\ \mathcal{S} & \end{pmatrix} = \begin{pmatrix} 0 & 1 \\ 1 & 0 \end{pmatrix} \otimes \begin{pmatrix} \mathbb{1} & \\ & -\mathbb{1} \end{pmatrix}, \tag{S42}$$

where $\mathcal{S}_H$ is the extended Hermitian chiral symmetry from $\mathcal{S}$. By construction, $\mathcal{H}_H$ enjoys an additional chiral symmetry $\Sigma$,

$$\{\Sigma, \mathcal{H}_H\} = 0, \quad \Sigma = \begin{pmatrix} 1 & \\ & -1 \end{pmatrix} \otimes \begin{pmatrix} \mathbb{1} & \\ & \mathbb{1} \end{pmatrix}. \tag{S43}$$

Note that the combination of two chiral operators $\mathcal{S}_H$ and $\Sigma$ ($\mathcal{S}_H\Sigma$) commutes with $\mathcal{H}_H$, i.e.,

$$[\mathcal{S}_H \Sigma, \mathcal{H}_H] = 0, \quad \mathcal{S}_H \Sigma = \begin{pmatrix} 0 & -1 \\ 1 & 0 \end{pmatrix} \otimes \begin{pmatrix} \mathbb{1} & \\ & -\mathbb{1} \end{pmatrix} = -i\tau_2 \otimes \sigma_3 \otimes \mathbb{1}. \tag{S44}$$

The combined operator $\Sigma \mathcal{S}_H$ has eigenvalues $\pm i$. Thus, we can block-diagonalize $\mathcal{H}_H$ with each block labeled by the eigenvalues $\pm i$, using the unitary matrix $U$ composed of eigenvectors of $\mathcal{S}_H \Sigma$. The resultant block-diagonalized Hamiltonian reads,

$$U \mathcal{H}_H U^{-1} = \begin{pmatrix} i\mathcal{H}_{\text{nh}} \mathcal{S} & \\ & -i\mathcal{H}_{\text{nh}} \mathcal{S} \end{pmatrix}, \tag{S45}$$

where $i\mathcal{H}_{\text{nh}} S$ is Hermitian,

$$(i\mathcal{H}_{\text{nh}} \mathcal{S})^\dagger = i\mathcal{H}_{\text{nh}} \mathcal{S}. \tag{S46}$$

Therefore, we can use the Hermitian Hamiltonian $i\mathcal{H}_{\text{nh}} S$ to capture the topology of the non-Hermitian point-gapped Hamiltonian $\mathcal{H}_{\text{nh}}$ with chiral symmetry $\mathcal{S}$.

## VI. NON-HERMITIAN GENERALIZED SPINLESS KANE-MELE MODEL

The non-Hermitian generalized spinless Kane-Mele model

$$\mathcal{H}_{\mathrm{KM}}(\mathbf{k}) = t\chi_1(\mathbf{k})\tau_0 \otimes \sigma_1 + t\chi_2(\mathbf{k})\tau_0 \otimes \sigma_2 - 2iJ_R\eta_1(\mathbf{k})\tau_0 \otimes \sigma_3 - 2iJ_I\eta_2(\mathbf{k})\tau_2 \otimes \sigma_0, \tag{S47}$$

in the main text has the chiral symmetry of

$$\mathcal{S}\mathcal{H}_{\mathrm{KM}}(\mathbf{k})^\dagger \mathcal{S}^{-1} = -\mathcal{H}_{\mathrm{KM}}(\mathbf{k}), \quad \mathcal{S} = \tau_0 \otimes \sigma_3. \tag{S48}$$

According to the discussion in the previous section, the topology of the non-Hermitian $\mathcal{H}_{\mathrm{KM}}(\mathbf{k})$ can be captured by the Hermitian Hamiltonian of

$$i\mathcal{H}_{\mathrm{KM}}(\mathbf{k})\mathcal{S} = t\chi_1(\mathbf{k})\tau_0 \otimes \sigma_2 - t\chi_2(\mathbf{k})\tau_0 \otimes \sigma_1 + 2J_R\eta_1(\mathbf{k})\tau_0 \otimes \sigma_0 + 2J_I\eta_2(\mathbf{k})\tau_2 \otimes \sigma_3, \tag{S49}$$

which satisfies the time reversal $T$ as

$$T(i\mathcal{H}_{\mathrm{KM}}(\mathbf{k})\mathcal{S})T^{-1} = i\mathcal{H}_{\mathrm{KM}}(-\mathbf{k})\mathcal{S}, \quad T = i\tau_2 \otimes \sigma_3 KI. \tag{S50}$$

Since $T^2 = -1$, the 2D Hermitian Hamiltonian $i\mathcal{H}_{\mathrm{KM}}(\mathbf{k})\mathcal{S}$ belongs to class AII and is characterized by the $\mathbb{Z}_2$ Kane-Mele invariant. Such a topological invariant can be computed by the Wilson loop spectrum, as discussed in the main text.

## VII. TOPOLOGICAL EDGE STATES OF THE NON-HERMITIAN GENERALIZED SPINLESS KANE-MELE MODEL

In this section, we discuss the topological features of the non-Hermitian generalized spinless Kane-Mele model, i.e., model (12) in the main text.

It is noteworthy that although the model looks like the Kane-Mele model, it is in the ramified symmetry class CII$^\dagger$, completely distinct from the class AII of the Kane-Mele model. For Hermitian systems, class CII has no topological phase in two dimensions. Thus, the $\mathbb{Z}_2$ classification for CII$^\dagger$ in two dimensions really stems from the non-Hermiticity.

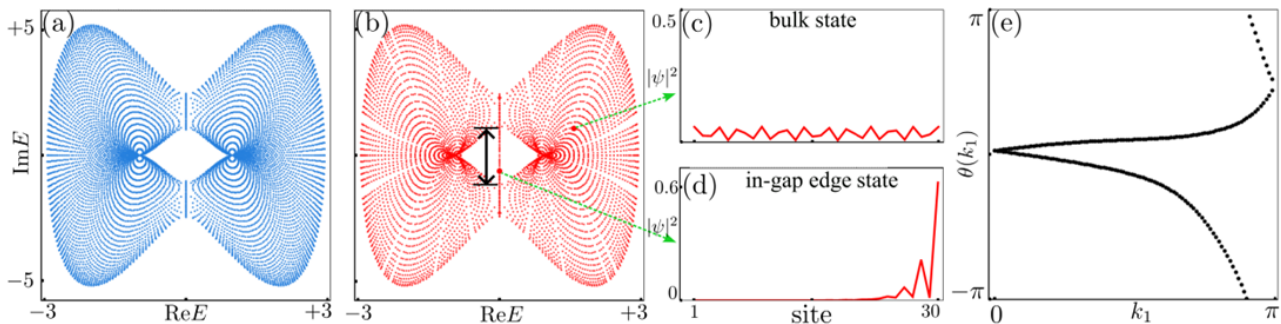

Figure S6. Energy spectra of the system under PBC (a), and OBC (b) in the $\mathbf{b}_1$ direction. In (b), the double arrow line in black marks the region of in-gap edge states. Wavefunction profiles for a randomly chosen bulk state (c) and in-gap state (d). (e) Wilson loop spectrum corresponds to the in-gap state of (d).

Let us take OBC in the $\mathbf{b}_1$ direction, i.e., the direction along the primitive vector $\mathbf{b}_1$ of the hexagonal lattice, and then calculate the energy spectrum. The OBC spectrum is plotted in Fig. S6(b), in comparison with the PBC spectrum in Fig. S6(a). We find that there is no appearance of the non-Hermitian skin effect in this case, which is consistent with the fact that OBC spectrum is almost the same as the PBC. Note that the non-Hermitian skin effect is signified by the stark difference between the two spectra.

Interestingly, by comparing the PBC spectrum and OBC spectrum, we find that in-gap states show up in OBC spectrum. These states reside within the point gap region, and thus, are dubbed in-gap states. They are highlighted by the double-arrow line in Fig. S6(b). We plot the wavefunction profiles of a randomly chosen bulk state in Fig. S6(c) and in-gap state in Fig. S6(d). While the bulk states are extended in the bulk, the in-gap states are localized at the boundary. These in-gap states and bulk states behave the same as topological edge sates and bulk states in

conventional Hermitian systems. Thus, the topological feature of this system could be topological edge states under OBC.

To identify these in-gap states as topological edge states, we further calculate the topological invariants corresponding to these in-gap states. Note that a reference energy $E_{\text{ref}}$ can be added to the extended Hermitian Hamiltonian,

$$\mathcal{H}_H(\mathbf{k}) = \begin{pmatrix} & \mathcal{H}_{\text{KM}}(\mathbf{k}) - E_{\text{ref}} \\ (\mathcal{H}_{\text{KM}}(\mathbf{k}) - E_{\text{ref}})^\dagger & \end{pmatrix}, \tag{S51}$$

for calculating the invariants, where $\mathcal{H}_{\text{KM}}(\mathbf{k})$ is Eq. (12) in the main text. Due to the symmetry constraints in class CII$^\dagger$, $E_{\text{ref}}$ can only take imaginary values. This is consistent with the results that the in-gap states are all located on the imaginary energy axis. For the randomly chosen in-gap state in in Fig. S6(d), we can set $E_{\text{ref}}$ in Eq. (S51) the same as the energy of this state. The topological invariant can then be obtained by the Wilson loop of Eq. (S51). As shown in Fig. S6(e), the Wilson loop crosses the reference line $\theta = \pi$ once, meaning the $\mathbb{Z}_2$ topological invariant is non-trivial. A similar procedure can be applied to all in-gap states to show that they correspond to non-trivial topological invariants. We note that these results are consistent with a recent preprint of Ref. [5] discussing the bulk-boundary correspondence of point-gap topological phases.


\end{document}